\documentclass[twoside,8pt]{article}
\usepackage{epsfig}
\usepackage{amsmath,amsfonts}

\def\e{\textrm{e}}

\newcommand{\be}{\begin{equation}}
\newcommand{\ee}{\end{equation}}
\newcommand{\bea}{\begin{eqnarray}}
\newcommand{\eea}{\end{eqnarray}}

\begin{document}

\title{Information and complexity measures in the  interface of a metal and a superconductor}

\author{Ch.C. Moustakidis and C.P. Panos\\
$^{}$ Department of Theoretical Physics, Aristotle University of
Thessaloniki, \\ 54124 Thessaloniki, Greece }

\maketitle

\begin{abstract}
Fisher information, Shannon information entropy  and Statistical  Complexity are calculated for the interface of a normal metal and a superconductor, as a function of the temperature for several materials. The order parameter $\Psi({\bf r})$ derived from the  Ginzburg-Landau theory is used as an input together with  experimental values of critical transition temperature $T_c$ and the superconducting coherence length $\xi_0$. Analytical expressions are obtained for  information and complexity measures. Thus $T_c$ is directly related in a simple way with disorder and complexity.
An analytical relation is found of the Fisher Information with the energy profile of superconductivity  i.e. the ratio of surface free energy and the bulk free energy.  We verify that a simple relation holds between Shannon and Fisher information  i.e. a decomposition of a global information quantity  (Shannon) in terms of two local ones (Fisher information), previously derived and verified for atoms and molecules by Liu  et al. Finally, we find analytical expressions for generalized information measures like the Tsallis entropy and Fisher information. We conclude that the proper value of the non-extensivity parameter $q\simeq 1$, in agreement with previous work using a different model, where $q\simeq 1.005$.
\\
\\
Keywords: Information Theory; Fisher Information; Shannon Entropy; Tsallis Entropy; Statistical Complexity; Superconductivity; Order Parameter.

%

%


\end{abstract}

\section{Introduction}
Fisher information \cite{Fisher-22} has two basic roles to play in
theory \cite{Frieden-book}. First, it is a measure of the ability
to estimate a parameter; this makes it a cornerstone of the
statistical field of study called parameter estimation. Second, it
is a measure of the state of disorder of a system or phenomenon, an important aspect of physical theory
\cite{Frieden-book}. Fisher's information measure (FIM) is defined for the simplest case of one-dimensional probability distribution $P(x)$ as
 a functional of $P(x)$ i.e.
\begin{equation}
{\cal I}_F\equiv \int P(x)\left( \frac{d {\rm ln}
P(x)}{ d x}\right)^2 {\rm d}x= \int
\frac{1}{P(x)}\left( \frac{d P(x)}{d x}\right)^2 {\rm d}x.  \label{Fish-1}
\end{equation}
${\cal I}_F$ can also  be written in terms of the so  called  Fisher information density $i(x)=\frac{1}{P(x)}\left( \frac{d P(x)}{d x}\right)^2$ i.e. ${\cal I}_F=\int  i(x) dx$.
The FIM, according to its definition, is an accounter of  the sharpness
of the probability density. A sharp and strongly localized probability
density gives rise to a larger value of Fisher information. Its
appealing features differ appreciably from other information
measures because of its {\it local} character, in contrast with
the {\it global} nature of several other functionals, such as the
Shannon~\cite{Shannon-48}, Tsallis~\cite{Tsallis-88,Tsallis-2009} and Renyi~\cite{Renyi-70} entropies. The local character of
Fisher information shows  an enhanced sensitivity to strong
changes, even over a very small-sized region in the domain of
definition, because  it is as a functional of the
distribution gradient.

Very interesting  applications of Fisher information have been made in
quantum systems
\cite{Garbaczewski-06,Szabo-08,Dehesa-09,Sen-07a,Borgoo-08,Sanudo-08,
Nagy-06,Sen-07b,Howard-08,Romera-08,Flores-017}
including atoms, molecules, nuclei, or in mathematical physics in
general
\cite{Frieden-010,Sun-010,Olivares-010,Hernando-09,Ubriaco-09,
Pennini-09,Feng-09,Crooks-07,Pennini-06,Curilef-05,Nettleton-03,Chimento-02,Frieden-02a,
Frieden-02b,Chimento-00,Romera-98,Liu-07,Liu-95,Sen-05,Amovilli-04,March-013,Grassi-98}.
However, in spite of extensive applications of information theory in quantum
many body systems, most of the them focus on the behavior of systems in a single phase.
Actually,  there are a limited  number of  information-theoretical applications to  systems that undergo    a phase transition (see for example~\cite{Prokopenko-011,Ma-09,Huang-16,Wang-14,Ye-16,Gleiser-15,Sowisnski-17}).   The main motivation of the present work is to extend the application of FIM in order to include systems  in  a phase transition by
employing the phenomenological theory of Ginzburg-Landau. More
precisely, we consider the order parameter $\Psi({\bf r})$ as the
basic ingredient of the FIM functional. In particular, we {\it replace}  the probability distribution $P(x)$ with the inhomogeneous distribution of the superconducting  phase defined in a proper way. In this  model,  the Fisher information measure is introduced in a convenient way, directly related with the characteristics of a phase transition. We focus on the case of the interface between a normal metal and a
superconductor. We find the dependence of the FIM on  specific properties of the  interface of superconductors including  the superconducting coherence length as well as the critical  transition temperature.

Next, we calculate the Shannon information  entropy~\cite{Shannon-48} and the  LMC (statistical) complexity~\cite{Shiner-99,Lopez-95}.  The dependence of these measures on the temperature is examined and interesting comments are made.
The question of generalized (non-extensive) information measures is also addressed. An analytical relationship between Shannon and Fisher information, previously proved and shown to hold for atoms and molecules, is demonstrated in the present case of the superconducting interface as well.

This letter is organized as follows. In section~2 we review briefly the Ginzburg-Landau theory for inhomogeneous systems. In  Section~ 3 we present  the Fisher information measure using a probability distribution related with the order parameter.  In   section~4 we calculate and discuss  some additional information and complexity  measures including generalized ones. In section~5 we demonstrate Liu's identity, while in section~6 we discuss the possibility of alternative probability distributions. Finally, section~7 contains our concluding remarks.


\section{Ginzburg-Landau theory for inhomogeneous systems}
The Ginzburg-Landau theory is a theory of second-order phase
transition, where one introduces an {\it order parameter}
$\Psi({\bf r})$ which is zero above the transition temperature
$T_c$,  but takes a finite value for $T<T_c$ and uses the symmetry
of the relevant Hamiltonian to restrict the form of the free
energy as a functional of $\Psi({\bf r})$ \cite{Leggett-2006}.
Following the discussion by Leggett \cite{Leggett-2006}, we assume that
$F[\Psi({\bf r}),T]$ is the space integral of a free energy
density ${\cal F}$ which is a function only of $\Psi({\bf r})$ and
its space derivatives and also their complex conjugates. The form of the free energy functional
is
\cite{Leggett-2006}
\begin{equation}
F_s(T)= \int {\cal F}_s [\Psi({\bf r}),T]{\rm d}{\bf r},
\label{F-1}
\end{equation}
\begin{equation}
{\cal F}_s [\Psi({\bf r}),T]\equiv {\cal F}_n
(T)+\alpha(T)|\Psi({\bf r})|^2+\frac{1}{2}\beta(T) |\Psi({\bf
r})|^4+\gamma(T)|\nabla \Psi({\bf r})|^2, \label{F-2}
\end{equation}
where ${\cal F}_s$ is  the {\it superconducting} free energy density
and ${\cal F}_n$ is the {\it normal-state} free energy density.  In
eq.~(\ref{F-2}) the normalization of the order parameter
$\Psi({\bf r})$ is arbitrary. One demands that $\Psi({\bf r})$
should be zero above $T_c$ and take a uniform non-zero value for
$T<T_c$ leading to the following equalities for the coefficients
$\alpha(T)$, $\beta(T)$ and $\gamma(T)$ \cite{Leggett-2006}
\begin{eqnarray}
&&\alpha(T)\cong \alpha_0(T-T_c) \nonumber \\
&&\beta(T) \cong \beta(T_c)\equiv \beta\\
&&\gamma(T)\cong \gamma(T_c)\equiv \gamma.  \nonumber \label{coef-1}
\end{eqnarray}
To obtain the total free energy we must integrate this over the
system \cite{Leggett-2006,Annett-04}
\begin{equation}
F_s(T)=F_n(T)+\int \left( \alpha(T)|\Psi({\bf
r})|^2+\frac{1}{2}\beta |\Psi({\bf r})|^4+\gamma|\nabla \Psi({\bf
r})|^2  \right) {\rm d}{\bf r}. \label{free-2}
\end{equation}
According to Eq.~(\ref{free-2}) the free energy is a functional of
the scalar functions
 $\Psi({\bf r})$ and  $\Psi^*({\bf r})$.
In order to find the order parameter $\Psi({\bf r})$ we must
minimize the total free energy of the system. The
condition for the minimum free energy is found by performing a
functional differentiation with respect to the above functions that
is to solve the following two equations
\begin{equation}
\frac{\delta F_s[\Psi({\bf r}),T]}{\delta \Psi({\bf r})}=0, \quad  \frac{\delta F_s[\Psi({\bf r}),T]}{\delta \Psi^*({\bf r})}=0.
\label{cod-1}
\end{equation}
The above conditions can be
satisfied only when $\Psi({\bf r})$ obeys
\begin{equation}
-\gamma\nabla^2\Psi({\bf r})+\left(\frac{}{} \alpha
+\beta|\Psi({\bf r})|^2 \right)\Psi({\bf r}) =0. \label{free-3}
\end{equation}
Equation (\ref{free-3}) has several applications including the
properties of the surfaces and interface of superconductors.
Following Refs.~\cite{Gennes-66,Annett-04} we consider a simple model for the
interface between a normal metal and a superconductor. The
interface lies in the $yz$ plane separating the normal metal in
the $x<0$ region from the superconductor in the $x>0$ region. On
the normal metal side of the interface the superconducting order
parameter $\Psi({\bf r})$ must be zero. Now, assuming that $\Psi({\bf
r})$ must be continuous, then the following one  dimensional  nonlinear type
Schr\"{o}dinger equation must be solved,
\begin{equation}
-\gamma\frac{{\rm d}^2\Psi(x)}{{\rm
d}x^2}+\alpha(T)\Psi(x)+\beta\Psi^3(x)=0 \label{Free-4}
\end{equation}
in the region $x>0$ with the boundary condition $\Psi(0)=0$,
eq.~(\ref{Free-4}) can be solved analytically with the result~\cite{Gennes-66}
\begin{equation}
\Psi(x)=\Psi_0\tanh\left(\frac{x}{\sqrt{2}\xi(T)}\right).
\label{sol-1}
\end{equation}
Fig.~1 sketches the  spatial variation of the order parameter $\Psi(x)$ at the interface between a normal and superconducting metal  in the more general case of the presence of a magnetic field. The effective penetration depth $\lambda_{{\rm eff}}$ of the magnetic field is also displayed (for more details see Ref.~\cite{Gennes-66}).
In Eq.~(\ref{sol-1}) $\Psi_0$
is the value of the order parameter in the bulk far from the
surface and the parameter $\xi(T)$ is defined as
\begin{equation}
\xi(T)=\left(\frac{\gamma}{a(T)} \right)^{1/2}=\left(\frac{\gamma}{\alpha_0}\right)^{1/2}\frac{1} {\sqrt{T_c-T}}.
\label{xi-1}
\end{equation}
Considering that $\xi(0)\equiv\xi_0$ is the value of $\xi$ for $T=0$ the above relation is rewritten as
\begin{equation}
\xi(T)=\xi_0\frac{1}{\sqrt{1-\frac{T}{T_c}}}, \quad \xi_0=\left(\frac{\gamma}{a_0T_c}   \right)^{1/2}.
\label{xi-1-b}
\end{equation}
The quantity $\xi$ has dimensions of length and is  known as the Ginzburg-Landau coherence  length or healing length.  The
physical significance of this length, in the condensed phase, is
that  it is a measure of the minimum distance over which one can
"bend" the order parameter either in magnitude or in phase, before
the bending energy becomes comparable to the condensation energy
\cite{Leggett-2006}. The coherence length plays for the superconducting  systems a role closely  analogous to the healing length in a dilute Bose condensate~\cite{Leggett-2006} and  arises in all problems related with inhomogeneous superconductors, including surfaces, interfaces, defects and vortices~\cite{Annett-04}.

\section{The Fisher information measure as a functional of the order parameter }
The next step is to construct a {\it bridge} to connect the order parameter $\Psi({\bf r})$ with the Fisher information measure. At this point it will be helpful to follow the related discussion of Leggett~\cite{Leggett-2006}. According to this  analysis Ginzburg and Landau guess that the order parameter $\Psi({\bf r})$ has the nature of a {\it macroscopic} wave function. Actually, according to BCS theory it is indeed (up to normalization) the center-of mass wave function of the Cooper pairs. It is characterized also {\it macroscopic} in the sense that it is that unique eigenfunction of the two-particle density matrix which is associated with a macroscopic eigenvalue~\cite{Leggett-2006}.

According to the above analysis one  can proceed one more step by considering that (since $\Psi({\bf r})$ is in general a complex quantity) the function $|\Psi({\bf r})|^2$ represents a probability distribution  for the Cooper pairs in a superconductors~\cite{Annett-04}. Moreover, in the present specific problem we consider  that a suitable choice of the probability distribution $P(x)$ involved in Eq.~(\ref{Fish-1}) is the following
\begin{equation}
P(x)=\frac{1}{\sqrt{2}\xi}\left(1-\left(\frac{\Psi(x)}{\Psi_0}\right)^2\right)  =\frac{1}{\sqrt{2}\xi}\left(1-\tanh^2\left(\frac{x}{\sqrt{2}\xi}\right)\right).
\label{px-1}
\end{equation}
The  probability distribution $P(x)$ as a function of the distance $x$ is displayed in Fig.~2(a). Since $\int_0^{\infty}|\Psi(x)|^2 dx$ diverges,  (the number of particles is not conserved in superconductivity) we employ an associated probability distribution $P(x)$ to describe the transition from metal to the superconducting phase (the interface region).
The proposed distribution firstly  ensures  the convergence of  the integral of $P(x)$  and secondly satisfies  the normalization condition  $\displaystyle \int_{x=0}^{\infty} P(x) dx=1$.   Moreover, $P(x)$ given by (\ref{px-1})  incorporates all the characteristics of the order parameter $\Psi(x)$.
Obviously,  $P(x)$ has the dimension of  inverse length and consequently the FIM has the dimension of  inverse length squared.  We comment on alternative, possible choices of $P(x)$ in section~6.

Now, the Fisher information measure, defined in Eq.~(\ref{Fish-1}), by employing the  probability distribution (\ref{px-1}), is given by
\begin{equation}
{\cal I}_F=\frac{4}{(\sqrt{2}\xi)^2}\int_0^{\infty} \left(\frac{\tanh(x)}{\cosh(x)}\right)^2 dx.
\label{IFint-1}
\end{equation}
The integral is easily calculated and gives the factor $1/3$. In total, the FIM  exhibits the following analytical and very simple relation  with the coherence length $\xi$
\begin{equation}
{\cal I}_F=\frac{2}{3}\frac{1}{\xi^2}.
\label{IF-GL-1}
\end{equation}
The above dependence is displayed in Fig.~2(b).  This result confirms both qualitatively  and quantitatively   the statement that FIM is a measure of the state of the disorder of a system. In particular, the smaller  $\xi$ (fast recovery of the order parameter to its bulk value), the higher is the value of the Fisher measure. Combining (\ref{xi-1-b}) and (\ref{IF-GL-1}), we obtain ${\cal I}_F$ as a function of $T/T_c$ (see Fig.~2(c) and also Fig.~3).

It is of interest to examine also the dependence of FIM on other physical quantities related with the phase transition. Thus from Eqs.~(\ref{xi-1}) and (\ref{IF-GL-1}) we get its temperature dependence:
\begin{equation}
{\cal I}_F=\frac{2\alpha_0}{3\gamma}(T-T_c).
\label{IF-T}
\end{equation}
The FIM attains its maximum value ${\cal I}_F^{max}=-2\alpha_0 T_C/3\gamma$ at temperature $T=0$, that is for $\xi=\xi_0$, while it becomes zero when the superconducting  phase disappears (for $T\geq T_c$). It is worth to point out that the FIM is connected with some experimental quantities specifically  the critical temperature $T_c$ as well as  the values of $\xi(0)=\xi_0$. In particular, the zero temperature value of $\xi$ in  BCS theory is  related with the size of a single Cooper pair~\cite{Annett-04}.   In Figs.~2(c) we display also the linear dependence of the ratio ${\cal I}_F/{\cal I}_F^{max}$  on  $T/T_c$.
Hence, in a way, the Fisher information is an alternative  measure of the extent of the  interface between  a normal metal and a superconductor.

Moreover, the Fisher information related with the free energy characterizes a superconductor. In particular, the surface free  energy ${\cal F}_{{\rm surf}}(T)$ is the difference between the condensation free energy $F_s(T)-F_n(T)$ defined in Eq.(\ref{free-2}) (a measurable quantity~\cite{Leggett-2006}) and the  bulk free energy density ${\cal F}_{{\rm bulk}}(T)$ which corresponds to an  uniformly  superconducting  medium in the region  from $x=0$ to $x=\infty$. It was found that \cite{Gennes-66,Annett-04}
\begin{equation}
{\cal F}_{{\rm surf}}(T)=\frac{4\sqrt{2}}{3}{\cal F}_{{\rm bulk}}(T)\xi(T).
\label{Esf}
\end{equation}
Now, by combining Eqs.(\ref{IF-GL-1}) and (\ref{Esf}) we obtain  the  relation
\begin{equation}
{\cal I}_F=\left(\frac{4}{3} \right)^3\left(\frac{{\cal F}_{{\rm surf}}}{{\cal F}_{{\rm bulk}}} \right)^{-2}.
\label{IF-Es}
\end{equation}
Eq.~(\ref{IF-Es}) exhibits  a simple relation between the FIM and the  energy {\it profile} of  superconductivity.
Obviously, the Fisher information is a measure  of the  surface energy (in units of the bulk energy).  In Fig.~3 we plot the Fisher measure as a function of temperature for four materials.  We observe  a strong dependence of ${\cal I}_F$ on $\xi_0$. In any case,   it is remarkable to connect directly a  theoretical quantity ${\cal I}_F$ with two experimentally measured quantities  $T_c$ and $\xi_0$.

\section{Shannon entropy and  statistical complexity. Generalized information measures}
It is instructive to calculate  the Shannon entropy~\cite{Shannon-48}, and LMC statistical  complexity~\cite{Shiner-99,Lopez-95}. In particular, the Shannon entropy  defined as
\begin{eqnarray}
S=\int s(x) dx=-\int P(x)\ln P(x) {\rm d}x,
\label{Shannon-1}
\end{eqnarray}
where $s(x)= -P(x)\ln P(x)$ is the Shannon information density. Using  the  probability distribution (\ref{px-1}), $S$ exhibits the following dependence on $\xi$
\begin{equation}
S=2+\ln\left(\frac{\xi}{2^{3/2}}  \right).
\label{Shannon-2}
\end{equation}
Relations (\ref{Shannon-2}) and (\ref{xi-1-b}) provide the temperature dependence of $S$, shown in Fig.~4.
$S$ is a measure of the information content stored in a quantum system described by $P(x)$. The units of $S$ are nats for a natural logarithm or bits, if the base of the logarithm is 2.

The disequilibrium $D$ defined as $D=\int \left[P(x)\right]^2 dx$ is given by
\begin{equation}
D=\frac{\sqrt{2}}{3\xi}.
\label{Disequil-1}
\end{equation}
$D$ is the disequilibrium of the system i.e.  the distance from its actual state to equilibrium. The quantity $D$ is an experimentally  measurable quantity i.e. in quantum chemistry~\cite{Hyman-78} known as  quantum self-similarity~\cite{Carbo-80,Borgoo-07,Angulo-07} or information energy~\cite{Onicescu-66} or linear entropy~\cite{Hall-07,Pennini-07}. The dependence of $D$ on $T$ is plotted in Fig.~5.

The {\it statistical} measure of complexity $C$  introduced by Lopez-Ruiz, Calbet and Mancini (LMC)~\cite{Lopez-95} employs the information entropy $S$ or information stored in a system and its distance $D$ to the equilibrium probability distribution as the two basic ingredients giving the correct asymptotic properties of a well-behaved measure of complexity. The LMC measure is easily computable (in contrast  with other definitions e.g. algorithmic complexity~\cite{Kolmogorov-65,Chaitin-66}
defined as the length of the shortest possible program necessary to reproduce a given object), defined as a product $C=S\cdot D$. As expected intuitively it vanishes for the two extreme cases of a perfect crystal (perfect order) and ideal gas (complete disorder).
An investigation of $C$ in  a quantum many-body system was carried out in atoms, for continuous electron distributions~\cite{Panos-07} and discrete ones~\cite{Panos-09}. An alternative definition of complexity is the SDL measure $\Gamma_{\alpha\beta}$~\cite{Shiner-99} (Shiner, Davison, Landsberg) defined and calculated in a similar way as $C$. It has been applied in atoms starting from~\cite{Chatz-05}. A welcome property of a definition of complexity might be the following: If one complicates the system by varying some of its parameters, and this leads to an increase of the adapted measure of complexity, then one could argue that this measure describes the complexity of the system properly. A detailed discussion of the physical meaning of $C$ can be found in \cite{Panos-07}, section 4.

It is noted that $C$ cannot be measured experimentally, but it is possible to calculate it, starting from a reasonable probabilistic definition (e.g. LMC~\cite{Lopez-95} or SDL~\cite{Shiner-99}) by using an information-theoretical  method, developed in previous work see e.g. \cite{Panos-07,Panos-09,Chatz-05}.

The LMC complexity  $C=S\cdot D$ is  a product of two {\it global} information quantities $S$ and $D$. It is appropriate to examine $\tilde{C}=S\cdot I_F$ as well which is defined as a product of one global quantity $S$ by a local one, specifically the Fisher information ${\cal I}_F$.
Finally, we find that the  statistical complexities  $C$ and $\tilde{C}$ are given by
\begin{equation}
C=\frac{\sqrt{2}}{3\xi}\left[2+\ln\left(\frac{\xi}{2^{3/2}}   \right)\right],
\label{Coml-1}
\end{equation}
\begin{equation}
\tilde{C}=\frac{2}{3\xi^2}\left[2+\ln\left(\frac{\xi}{2^{3/2}}   \right)\right].
\label{Coml-2}
\end{equation}
The dependence of $C$ on $T$ is seen in Fig.~6(a) while $\tilde{C}$ is shown in Fig.~6(b).  The calculated values of ${\cal I}_F$, $C$ and $D$, follow the same trend as a function of $Z$ compared with experimental  $T_c$ (see Fig.~7). The only exception is the inverse behavior of $S$. Although $S$ and $D$ are reciprocal, their product (LMC complexity) $C=S\cdot D$ shows a trend similar to $T_c$. This is expected according to physical intuition: Larger (smaller) values of $T_c$ correspond to larger (smaller) complexity $C$, $D$, ${\cal I}_F$. In the present work we find that this holds for the interface metal-superconductor.

It is worth to extend also our study by including a generalization of the Shannon Entropy and Fisher measure in the framework of the nonextensive statistical mechanics.  This is based on the generalization of the definition of the logarithm according to the expression~\cite{Tsallis-88}
\begin{equation}
\ln_q x\equiv \frac{x^{1-q}-1}{1-q}
\label{NE-log-1}
\end{equation}
In view of the above definition,   the Fisher information  ${\cal I}_{q}$ and the Tsallis entropy $T_q$ (a generalization of the Shannon entropy) are defined as \begin{equation}
{\cal I}_{q}\equiv \int P(x)\left( \frac{d {\rm ln}_q P(x)}{d
x}\right)^2 {\rm d}x=\int \left[P(x)\right]^{1-2q}\left(\frac{dP(x)}{d x}  \right)^2 dx
\label{IF-q-1}
\end{equation}
and
\begin{equation}
T_{q}\equiv \int P(x)\ln_q \left(\frac{1}{P(x)}\right) dx=\frac{1}{q-1}\left(1-\int \left[P(x)\right]^q dx  \right)
\label{IS-q-1}
\end{equation}
Our results are
\begin{eqnarray}
{\cal I}_{q}&=&\frac{2^{-3+q}\xi^{2q-4}}{(2q-5)(2q-3)\Gamma\left(\frac{9}{2}-2q\right)}\nonumber \\
&\times&\left(-\frac{\sqrt{\pi}(q-1)\Gamma(6-2q)}{q-2}+2\ \Gamma\left(\frac{9}{2}-2q\right)\left[\frac{}{}5-2q+(3-2q) {_2F_1}\left(1,-2+2q,6-2q,-1\right)\right]  \right)
\label{IF-q-2}
\end{eqnarray}
and
\begin{equation}
T_{q}=\frac{1}{q-1}\left(1-\frac{\Gamma(q)}{\Gamma(1+q)}2^{\frac{1}{2}(3q-1)}\xi^{1-q} {_2F_1}\left(q,2q,1+q,-1\right)\right)
\label{IS-q-2}
\end{equation}
 where $\Gamma(x)$ is the usual  gamma  function and  and $_2F_1\left(a,b,c,x\right)$ is the corresponding  hypergeometric function.
The statistical complexity, in the framework of the nonextensive statistical mechanics,  is given  by
\begin{equation}
C_q=D_q\cdot T_q=\frac{\sqrt{2}}{3\xi}  \frac{1}{q-1}\left(1-\frac{\Gamma(q)}{\Gamma(1+q)}2^{\frac{1}{2}(3q-1)}\xi^{1-q} {_2F_1}\left(q,2q,1+q,-1\right)\right).
\label{Cq-1}
\end{equation}
It is noted  that  Eqs.~(\ref{IF-q-2}) and (\ref{IS-q-2}) for $q\rightarrow 1$ converge to the corresponding ones~(\ref{IF-GL-1}) and (\ref{Shannon-2}) respectively.
In Fig.~9 we display the Tsallis entropy $T_q$  as a function of temperature for Sn  for various values of q.
Previously, $T_q$ has been employed in more sophisticated attempts to estimate an optimal value of $q (q\simeq 1)$ in superconductivity~\cite{Lizardo-01,Lizardo-02}. It was found that small deviations with respect to $q=1$ i.e. $q\simeq 1.005$ provide better agreement with experimental results. A value of $q\simeq 1$ can also be inferred with the present model from  Fig.~9 by inspection. We present only the plot of $T_q$, while the other quantities $I_F$ and $C_q$ lead to the  same approximate conclusion about $q$.

\section{Analytical relation between Shannon  and  Fisher information densities }
Under the condition, for  systems such as atoms and molecules whose asymptotic density decays strongly, Liu et al.~\cite{Liu-07,Liu-95} obtained a specific relationship between their respective densities, $s({\bf r})$, $\rho({\bf r})$, $i_F({\bf r})$, $i_F'({\bf r})$  and consequently of the integral forms of $S$ and ${\cal I}_F$
\begin{equation}
s({\bf r})=-\rho({\bf r})+\frac{1}{4\pi}\int\frac{i_F({\bf r}')}{|{\bf r}-{\bf r}'|}d{\bf r}'-\frac{1}{4\pi}\int\frac{i_F'({\bf r}')}{|{\bf r}-{\bf r}'|}d{\bf r}'
\label{sx-1}
\end{equation}
Integration of both sides gives
\begin{equation}
S[\rho]=-N+\frac{1}{4\pi}\int\int\frac{i_F({\bf r}')}{|{\bf r}-{\bf r}'|}d{\bf r}'d{\bf r}-\frac{1}{4\pi}\int\int\frac{i_F'({\bf r}')}{|{\bf r}-{\bf r}'|}d{\bf r}'d{\bf r}.
\label{sx-2}
\end{equation}
The above equations are identities. Here
\begin{equation}
i_F({\bf r})=\frac{{\bf \nabla}\rho({\bf r})\cdot {\bf \nabla}\rho({\bf r})}{\rho({\bf r})}
\label{Iifr-1}
\end{equation}
and
\begin{equation}
i_F'({\bf r})=-{\bf \nabla}^2\rho({\bf r})\ln\rho({\bf r}) d{\bf r}.
\label{Iifr-2}
\end{equation}
We  also have
\begin{equation}
I_F=\int i_F({\bf r}) d{\bf r}
\label{IFr-1}
\end{equation}
\begin{equation}
I_F'=\int i_F'({\bf r}) d{\bf r}.
\label{IFr-2}
\end{equation}
The densities $i_F({\bf r})$, $i_F'({\bf r})$ are different functions of the vector ${\bf r}$, but their integrals are equal, as expected. For a strongly decaying local quantity $q({\bf r})$ the authors of Ref.~\cite{Liu-07,Liu-95} state that $q({\bf r})$ should decay faster than $1/{\bf r}$ and its derivative faster than  $1/{\bf r}^2$ as a requirement for their identities.

Liu's identity can be written in compact form:
\begin{equation}
S=-N+I_{F}+I_{F}'
\label{Liu-1}
\end{equation}
quite a remarkable  decomposition of a global information quantity $S$ in terms of (two) local ones,  $I_F$ and $I_F'$.

In our present work we show that an analogous relation holds for the special case of the  one-dimensional density  $P(x)$ and verify it in superconductivity. Again, we checked that $P(x)$ satisfies the same criterion of strong decay for $x \rightarrow \infty$. In particular we start from the identity

\begin{equation}
q(x)\equiv q(0)+\int_{0}^x\frac{d^2q(x')}{dx'^2}(x-x')dx'
\label{id-1}
\end{equation}
which  holds under  the condition
\[\left(\frac{dq(x)}{dx}\right)_{x=0}=0 . \]
Now, we define the local Shannon information density
\begin{equation}
s(x)=-P(x)\ln P(x)
\label{sx-1}
\end{equation}
which obviously leads to the Shannon entropy
\begin{equation}
S=\int_{0}^{\infty} s(x) dx=-\int_{0}^{\infty}P(x)\ln P(x)dx.
\label{shan-1}
\end{equation}
Now, according to the identity~(\ref{id-1}) we have also
\begin{equation}
s(x)=s(0)+\int_{0}^x\frac{d^2s(x')}{dx'^2}(x-x')dx'
\label{id-2}
\end{equation}
and using Eq.~(\ref{sx-1}) we obtain
\begin{eqnarray}
s(x)&=&s(0)-P(x)+P(0)-\int_{0}^x\frac{d^2P(x')}{dx'^2}\ln P(x')(x-x')dx'-
\int_{0}^x\frac{1}{P(x')}\left(\frac{dP(x')}{dx'}\right)^2(x-x')dx'\nonumber\\
&=&-P(x)+P(0)-\int_{0}^x\frac{1}{P(x')}\left(\frac{dP(x')}{dx'}\right)^2(x-x')dx'\nonumber \\
&+&s(0)-\int_{0}^x\frac{d^2P(x')}{dx'^2}\ln P(x')(x-x')dx'=-P(x)+I_{1f}(x)+I_{2f}(x)
\label{sx-2}
\end{eqnarray}
where
\begin{equation}
I_{1f}(x)=P(0)-\int_{0}^x\frac{1}{P(x')}\left(\frac{dP(x')}{dx'}\right)^2(x-x')dx'
\label{Ifx-1}
\end{equation}
\begin{equation}
I_{2f}(x)=s(0)-\int_{0}^x\frac{d^2P(x')}{dx'^2}\ln P(x')(x-x')dx'
\label{Ifx-2}
\end{equation}
Finally, integrating Eq.~(\ref{sx-2}) over  $x$ we get
\begin{equation}
S=-1+I_{F1}+I_{F2}
\label{shan-2}
\end{equation}
where
\begin{equation}
I_{1F}=\int_{0}^{\infty}I_{1f}(x) dx
\label{1f}
\end{equation}
\begin{equation}
I_{2F}=\int_{0}^{\infty}I_{2f}(x) dx
\label{2f}
\end{equation}
Equation~(\ref{shan-2}) is analogus with the Liu identity~(\ref{Liu-1}) for $N=1$, rewritten for the one dimensional case.

\begin{table}[h]
\begin{center}
\caption{ Experimental values of the coherence length $\xi_0$~\cite{Meser}, the Shannon information entropy $S$, the terms  $I_{1F}$   and  $I_{2F}$  and the sum $-1+I_{1F}+I_{2F}$ in order to confirm relation~(\ref{shan-2}).}
 \label{t:1}
\vspace{0.5cm}
\begin{tabular}{|c|c|c|c|c|c|c|}
\hline
Atom     & Z    &    $\xi_0$ (nm)  &   $S$   &   $I_{1F}$   & $I_{2F}$ & $-1+I_{1F}+I_{2F}$    \\
\hline
Al       & 13   &  1600   &     8.338   &  -13.337  & 22.675    &  8.338 \\
\hline
Nb       &  41    &  38 &     4.598     &   -828.154          &  833.752              & 4.598  \\
\hline
In        & 49   &  360     &   6.846      &  -207.527    &     215.374   &  6.846 \\
\hline
Sn      &  50   &  230     &     6.398      & -551.682      &  559.080     & 6.398 \\
\hline
Ga     & 64      &  760     &   7.594     &  -26.265    &   34.859    &   7.594 \\
\hline
Ta     & 73    &  93     &  5.493        & -288.979   & 295.471   &5.493 \\
\hline
Pb     & 82      &  83  &   5.379    &  -370.474    & 376.853   & 5.379\\
\hline
 \end{tabular}
\end{center}
\end{table}

\section{Alternative probability distributions}
The question naturally arises if instead of $P(x)$, relation  (\ref{px-1}), we may consider different, alternative definitions, repeat all our calculations and examine the consequence on our results, both qualitatively and quantitatively. First we see that the integral
\[
\int_{0}^{\infty} \left|\frac{\Psi(x)}{\Psi_0}   \right|^2 dx=\int_{0}^{\infty} \tanh^2\left(\frac{x}{\sqrt{2}\xi}\right) dx\]
diverges.
We may try a cutoff, i.e. integrate from $0$ to $n\xi$, where $n=1,2,3,\dots$. We note that it is not possible to know in advance or a priori the correct upper limit if such a limit does exist. We see that for $n=5$, $x=5\xi$:
\[\tanh\left(\frac{x}{\sqrt{2}\xi}\right)\simeq 0.9983   \]
for any material. Hence we define
\begin{equation}
\tilde{P}(x)={\cal N}\left(\frac{\Psi(x)}{\Psi_0}  \right)^2 ={\cal N}\tanh^2\left(\frac{x}{\sqrt{2}\xi}\right).
\label{px-1-new-2}
\end{equation}
The normalization condition $\int_{0}^{n\xi}\tilde{P}(x) dx=1$ leads to
\begin{equation}
{\cal N}=\left[\sqrt{2}\xi \left(\frac{n}{\sqrt{2}}-\tanh\left(\frac{n}{\sqrt{2}}\right)\right)   \right]^{-1}
\label{norm-new-2}
\end{equation}
The Shannon entropy and the Fisher information take the analytical forms
\begin{eqnarray}
S&=&-\ln({\cal N})-{\cal N}\xi\left[4n \ {\rm arc coth}\left(\e^{\sqrt{2}n}\right)+n\ln\left[\tanh^2\left(\frac{n}{\sqrt{2}}\right)  \right]-\sqrt{2}\ {\rm Li}_2\left(-\e^{-\sqrt{2}n}\right)    \right. \nonumber \\
&+& \left.\sqrt{2}\ {\rm Li}_{2}\left(\e^{-\sqrt{2}n}\right)-\sqrt{2}\left(-2+\ln\left[\tanh^2\left(\frac{n}{\sqrt{2}}\right)  \right]    \right)\tanh\left(\frac{n}{\sqrt{2}}\right)-\frac{\pi^2}{2\sqrt{2}}     \right]
\label{S-new}
\end{eqnarray}
\begin{equation}
I_F={\cal N}\frac{\sqrt{2}}{3\xi}{\rm sech}^3\left(\frac{n}{\sqrt{2}}\right)\left[3\sinh\left(\frac{n}{\sqrt{2}}\right)+\sinh\left(\frac{3n}{\sqrt{2}}\right)   \right]
\label{Fis-new-2}
\end{equation}
where ${\rm Li}_n(x)$  is the dilogarithm (or Spence's function) which is defined for $ |x|\leq 1$ either a) by a power series in $x$ i.e. $ {\rm Li}_2(x)=\sum_{k=1}^{\infty}\frac{x^k}{k^2}$ or b) by the integral i.e. $ {\rm Li}_2(x)=-\int_0^x\ln(1-u)/u du$~\cite{Lewin-81}.
We choose $n=5$ (integration from $0$ to $5\xi$), we calculate again  $S$, ${\cal I}_F$, $D$ and $C$ and present our results for various materials in Table~3 and plot in Fig.~8. We see that the absolute values of the relevant quantities depend on the choice of $n$, as expected and are different than Table~2. However, the qualitative trend of our calculated  quantities as function of $Z$ is the same that is the zig zag pattern in Figs~7 and 8. In addition, we mention that the absolute values of information quantities are not important, since $\ln\xi$ is known up to an additive constant $c$ (see relation~(\ref{Shannon-2})).  For example, if one wishes to change or eliminate units of $\xi$ divides $\xi$ inside the logarithm  $\xi$ by $c$. Accordingly  Fisher information is known up to a multiplicative  factor (see relation~(\ref{IF-GL-1})).  Thus, we select  to employ our definition $P(x)$ of relation (\ref{px-1}), in order to avoid the ambiguity of a cutoff. This choice may also be justified by Occam's razor, though falsifiability is the ultimate criterion of the quality of a model in physical science. An additional merit of $P(x)$ is that it enables us to use an infinite integration interval leading to a direct demonstration of Liu's identity. The fact that for two different distributions $P(x)$ and $\tilde{P}(x)$ we get the same behavior corroborates the general property that the normalization of $\Psi(x)$ is arbitrary. It is enough to carry out consistent calculations for information and complexity.

\section{Concluding remarks}
Concluding, we note that the  Fisher information is a qualitative as well as a  quantitative measure of the superconducting  phase  between  a normal metal and a superconductor. It shows a simple analytical  dependence on experimentally measurable quantities like  the correlation length $\xi$ and consequently  the critical temperature $T_c$.  Some useful insights have been obtained by calculating  a few  information and complexity  measures.

Our formalism allows us to verify an identity derived previously by Liu et al.~\cite{Liu-07,Liu-95} shown to hold for atoms and molecules. Thus we are able  to decompose  a global information measure (Shannon) in terms of two local ones  (Fisher) for the metal-superconductor interface. Liu's identity is an example of a relationship between Fisher information (local) with Shannon entropy (global) for a specific class of continuous probability distributions (asymptotic behavior). In fact, Fisher information may be considered as the limiting form of many different measures of information or in the words of B. Roy Frieden (Ref.~\cite{Frieden-book} p.39) it is a kind of "mother" information. To justisfy this statement one can show that ${\cal I}_F$ is the cross-entropy between a $P(x)$ and its infinitesimally shifted version $P(x+\Delta x)$. In a sense, ${\cal I}_F$ more generally results as a "cross-information" between $P(x)$ and $P(x+\Delta x)$ for a host of different types of information measures (Ref.~\cite{Caianiello-92}).

A planned future work will extend our study to include other phases of superconductors in connection with   the Fisher information measure (including  surfaces, defects, vortices e.t.c). A final comment seems appropriate, quoting \cite{Panos-07}. It is difficult to quantify complexity, a context dependent and multi-faceted concept. Choosing a pragmatic approach, we employ as a starting point a definition of complexity based on a probabilistic description of a quantum system, in the present case  a metal-superconductor interface.


\section*{Acnowledgements}
One of the authors (Ch.C.M.) would like to thank Dr. S. Tserkis for useful discussions in the early stage of the present work.


\begin{table}[h]
\begin{center}
\caption{ Experimental values of the coherence length $\xi_0$~\cite{Meser} and the critical temperature $T_c$~\cite{Crit} for various materials in comparison with calculated information and complexity  measures including the Shannon entropy $S$, the Fisher measure $I_F$, the Disequilibrium $D$ and the LMC Complexity $C$. The probability distribution $P(x)$ is used, relation  (\ref{px-1}).}
 \label{t:1}
\vspace{0.5cm}
\begin{tabular}{|c|c|c|c|c|c|c|c|}
\hline
Atom     & Z&    $\xi_0$ (nm) &   $T_c$($ ^0$K)& S & $I_F$ ($\times 10^{-5}$) & D($\times 10^{-4}$) & C($\times 10^{-3}$)    \\
\hline
Al       & 13   &  1600   &     1.175   & 8.338   & 0.0260    & 2.946    &   2.471  \\
\hline
Nb       &  41    &  38 &     9.25     & 4.598   & 46.168   &  124.054    & 5.704 \\
\hline
In        & 49   &  360     &     3.41   & 6.846   &  0.514    & 13.095    & 8.965 \\
\hline
Sn      &  50   &  230 &    3.72       &  6.398  &  1.260  &  20.496  & 13.113  \\
\hline
Ga     & 64      &  760     &   1.083     &  7.594  & 0.115   & 6.200   &  4.708   \\
\hline
Ta     & 73    &  93     &      4.47    &  5.493  & 7.700   & 50.680   & 27.840\\
\hline
Pb     & 82      &  83  &    7.2   &  5.379  & 9.670   & 56.790  & 30.547\\
\hline
 \end{tabular}
\end{center}
\end{table}



\begin{table}[h]
\begin{center}
\caption{ The same as in Table~2, using the probability distribution  $\tilde{P}(x)$, relation~(\ref{px-1-new-2}).}
 \label{t:1}
\vspace{0.5cm}
\begin{tabular}{|c|c|c|c|c|c|c|c|}
\hline
Atom     & Z&    $\xi_0$ (nm) &   $T_c$($ ^0$K)& S & $I_F$ ($\times 10^{-5}$)  & D ($\times 10^{-4}$) & C($\times 10^{-3}$)   \\
\hline
Al       & 13   &  1600   &     1.175   & 8.839   & 0.0205    & 1.514    &   1.338  \\
\hline
Nb       &  41    &  38 &     9.25     & 5.099   & 36.400   &  63.750    & 32.506 \\
\hline
In        & 49   &  360     &     3.41   & 7.347   &  0.405    & 6.730    & 4.945 \\
\hline
Sn      &  50   &  230 &    3.72       &  6.899  &  0.993  &  10.533  & 7.267  \\
\hline
Ga     & 64      &  760     &   1.083     &  8.095  & 0.091   & 3.188   &  2.581   \\
\hline
Ta     & 73    &  93     &      4.47    &  5.994  & 6.076   & 26.050   & 15.614\\
\hline
Pb     & 82      &  83  &    7.2   &  5.880  & 7.628   & 29.188  & 17.163\\
\hline
 \end{tabular}
\end{center}
\end{table}

\begin{figure}
\centering
\includegraphics[height=8.1cm,width=10cm]{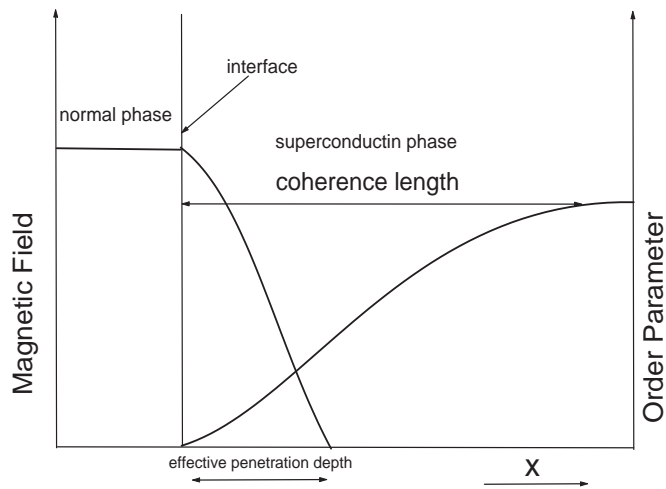}\
\caption{A schematic picture for the spatial variation of the order parameter $\Psi(x)$ at the interface between a normal and superconducting metal  in the presence of a magnetic field. The effective penetration depth $\lambda_{{\rm eff}}$ is also displayed. For more details see text and  Ref.~\cite{Gennes-66}.    } \label{picture}
\end{figure}
\begin{figure}
\centering
\includegraphics[height=5.5cm,width=5.7cm]{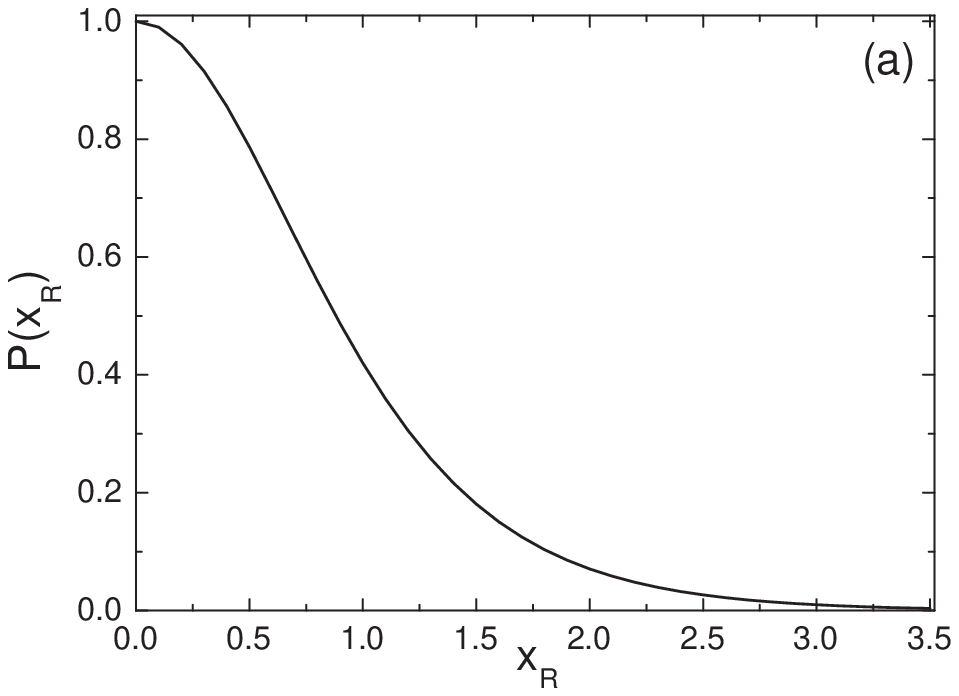}\
\includegraphics[height=5.5cm,width=5.7cm]{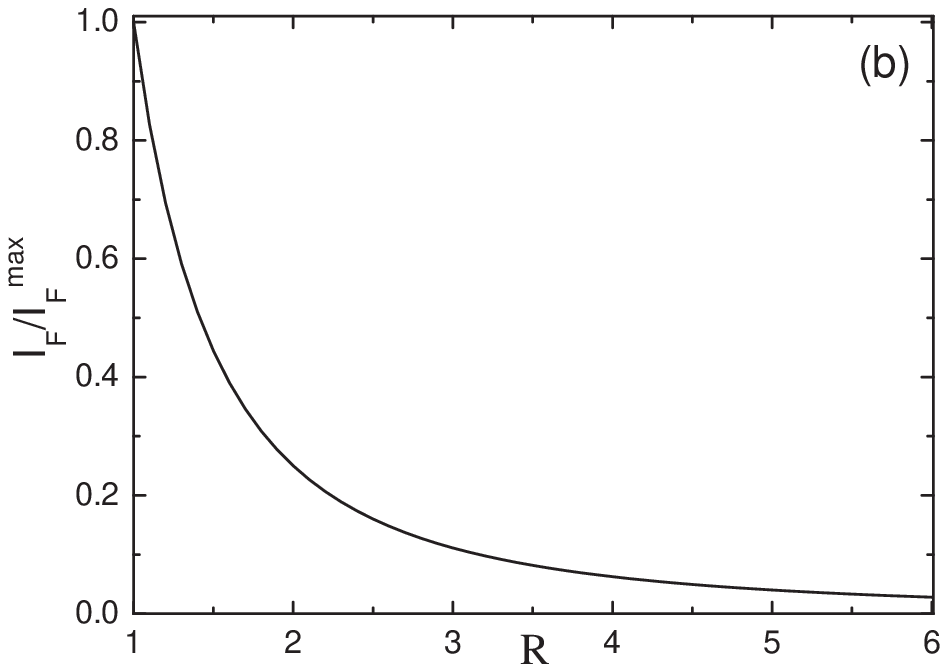}\
\includegraphics[height=5.5cm,width=5.7cm]{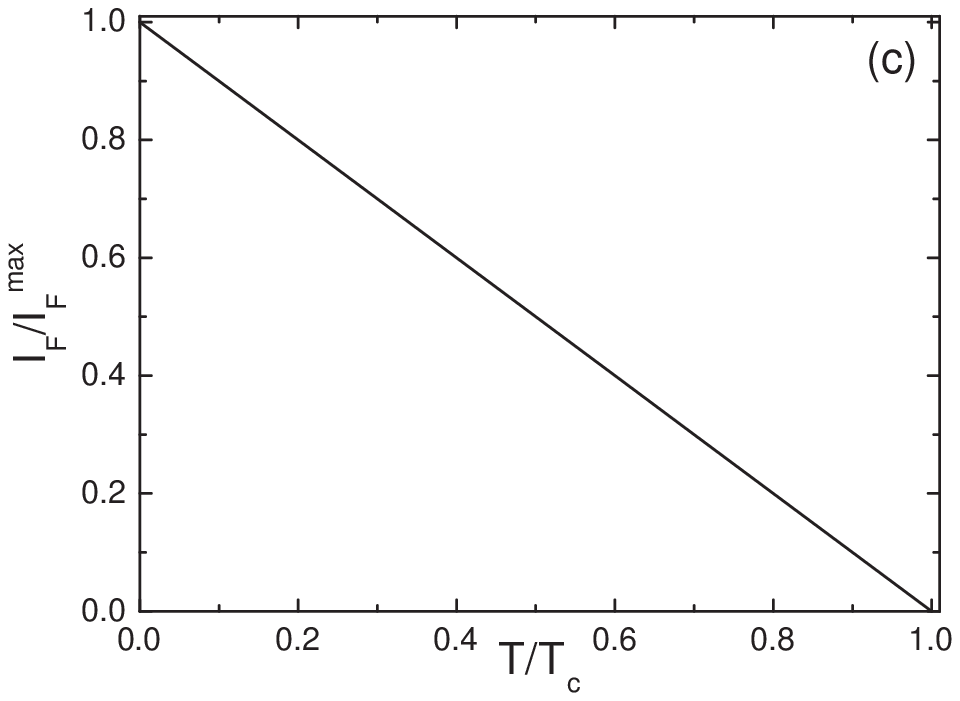}\
\caption{(a) The probability distribution function $P(x_R)=\sqrt{2}\xi P(x)$ as a function of the distance ratio $x_R=x/\sqrt{2}\xi$,  (b)  the Fisher's information measure $I_F$ (in units of $I_F^{max}$)   as a function of the coherence length ratio $R=\xi/\xi_0$ and (c) of the temperature ratio $T/T_c$.   }
\label{If-xi-T}
\end{figure}

\begin{figure}
\centering
\includegraphics[height=7.5cm,width=7.5cm]{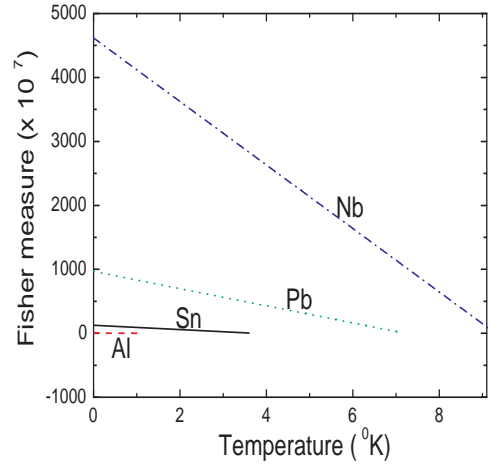}\
\caption{The Fisher measure as a function of temperature for four materials.}
\label{If-xi-T}
\end{figure}
\begin{figure}
\centering
\includegraphics[height=7.5cm,width=7.5cm]{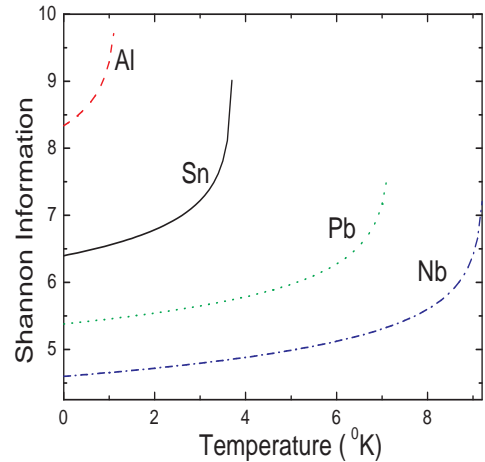}\
\caption{Temperature dependence of the Shannon information $S$ for four materials.}
\label{If-xi-T}
\end{figure}

\begin{figure}
\centering
\includegraphics[height=7.5cm,width=7.5cm]{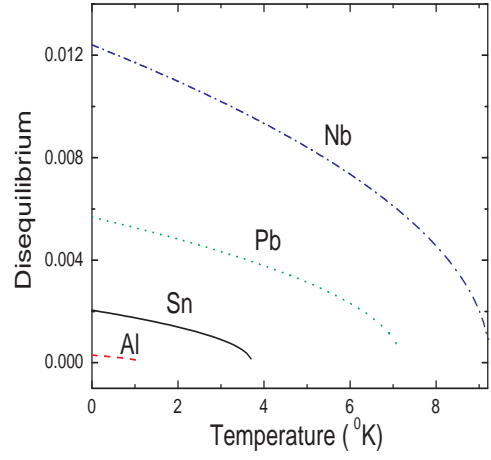}\
\caption{Temperature dependence of the Disequilibrium $D$ for four materials.}
\label{If-xi-T}
\end{figure}

\begin{figure}
\centering
\includegraphics[height=7.5cm,width=7.5cm]{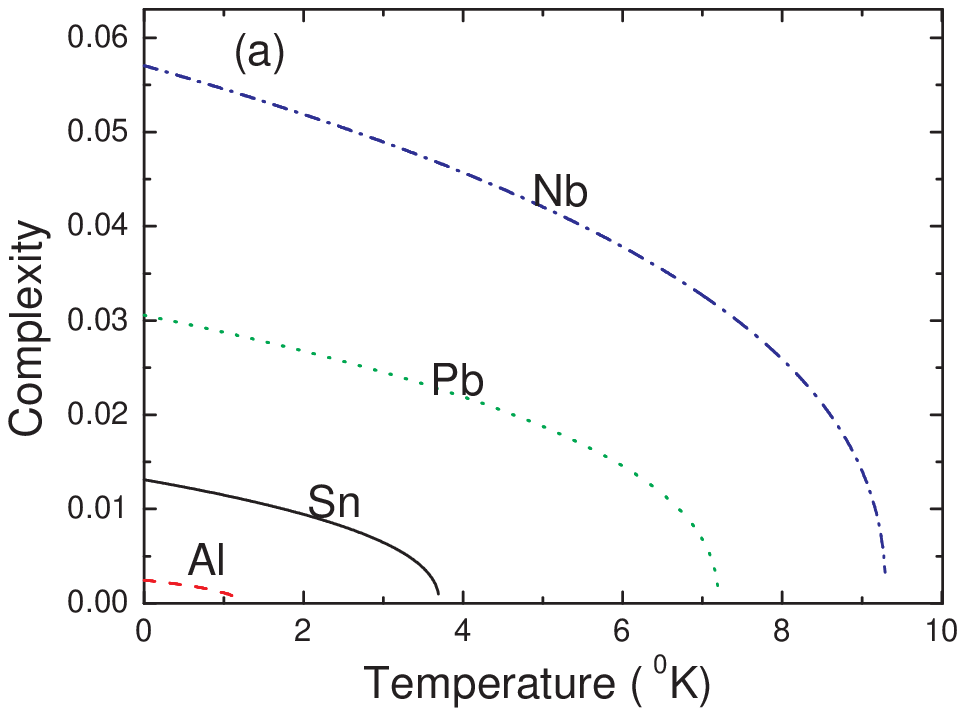}\
\includegraphics[height=7.5cm,width=7.5cm]{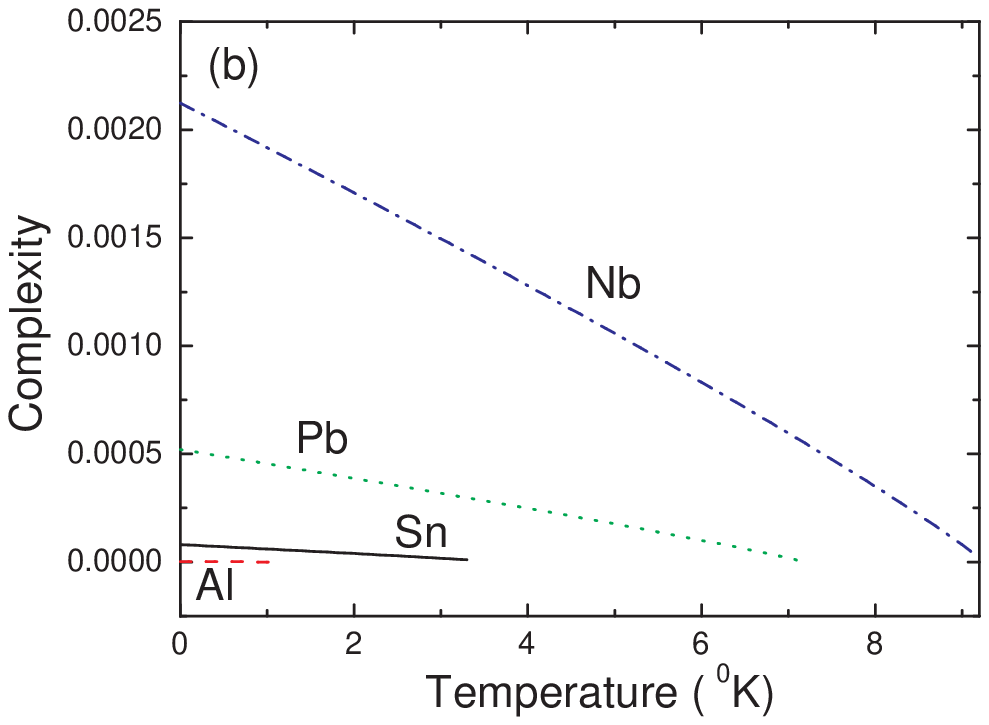}\
\caption{a) The LMC  complexity $C=S\cdot D$ as a function of temperature for four materials. (b) The complexity $\tilde{C}=S\cdot {\cal I}_F$ as a function of temperature for four materials.    }
\label{If-xi-T}
\end{figure}

\begin{figure}
\centering
\includegraphics[height=8cm,width=10cm]{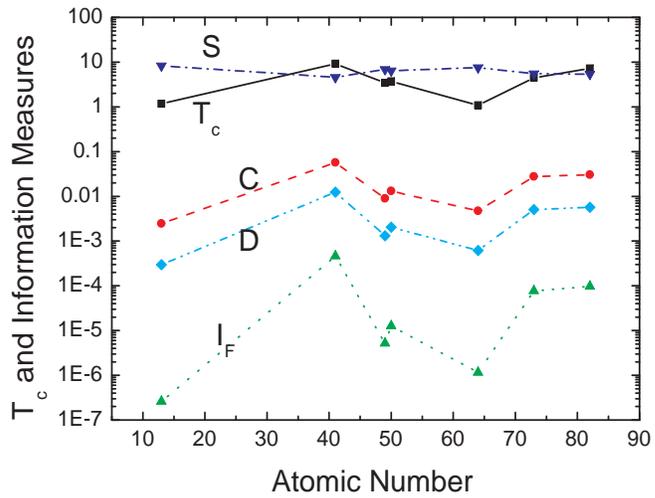}\
\caption{The critical temperature $T_c$ and the calculated values of  LMC complexity $C$, the Fisher measure ${\cal I}_F$, the Shannon entropy $S$ and the disequilibrium $D$  as  functions of the atomic number $Z$ for various materials (see also Table~2). We use the probability distribution $P(x)$, relation  (\ref{px-1}).  }
\label{If-xi-T}
\end{figure}

\begin{figure}
\centering
\includegraphics[height=8cm,width=10cm]{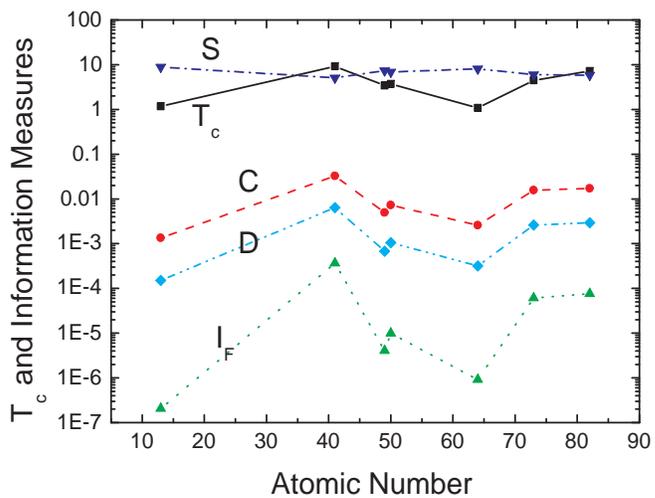}\
\caption{The same as in Fig.~7, but here we use the  probability distribution  $\tilde{P}(x)$, relation~(\ref{px-1-new-2}) (see also Table~3). }
\label{If-xi-T}
\end{figure}

\begin{figure}
\centering
\includegraphics[height=8cm,width=10cm]{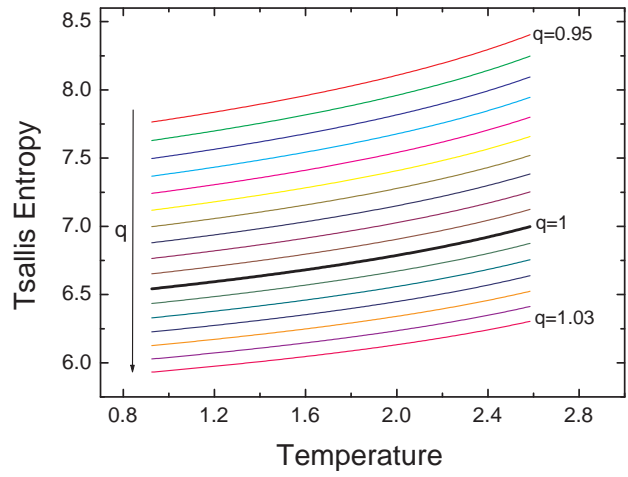}\
\caption{The Tsallis entropy as a function of the temperature, for Sn and   $0.95 <q<1.03$. }
\label{If-xi-T}
\end{figure}

\end{document}